\documentclass{article}

\usepackage{arxiv}

\usepackage[utf8]{inputenc} 
\usepackage[T1]{fontenc}    
\usepackage{hyperref}       
\usepackage{url}            
\usepackage{booktabs}       
\usepackage{amsfonts}       
\usepackage{nicefrac}       
\usepackage{microtype}      
\usepackage{lipsum}
\usepackage{graphicx}
\graphicspath{ {./images/} }

\title{A Comparison of Amputee and Able-Bodied Inter-Subject Variability in Myoelectric Control}

\author{
 Evan Campbell \\
  Institute of Biomedical Engineering\\
  University of New Brunswick\\
  \texttt{evan.campbell1@unb.ca} \\
   \And
    Jason Chang \\
  Institute of Biomedical Engineering\\
  University of New Brunswick\\
  \texttt{jason.chang@unb.ca} \\
   \And
 Angkoon Phinyomark \\
  Institute of Biomedical Engineering\\
  University of New Brunswick\\
  \texttt{aphinyom@unb.ca} \\
  \And
  Erik Scheme \\
  Institute of Biomedical Engineering\\
  University of New Brunswick\\
  \texttt{escheme@unb.ca} \\ \\
}

\begin{document}
\maketitle
\begin{abstract}
Despite decades of research and development of pattern recognition approaches, the clinical usability of myoelectric-controlled prostheses is still limited. One of the main issues is the high inter-subject variability that necessitates long and frequent user-specific training. Cross-user models present an opportunity  to  improve  clinical  viability  of  myoelectric control   systems by   leveraging   existing   data   to shorten training.However, due to the difficulty of obtaining large sets of  data  from  amputee  populations,  data  from  intact-limbed subjects are  often  supplemented when building  cross-user models; which may not translate well to clinical usability. In this preliminary study, the differences between intact-limbed and  amputee cross-user electromyography (EMG) patterns were  examined.Previously  collected EMG data  from  20 intact-limbed  and  10  amputee  subjects  for different wrist, finger,  and  grasping gestures  were  analysed.  Results using unsupervised    clustering showed    that amputees    were consistently grouped into  a  different  cluster  than intact-limbed  subjects and  that  additional  clustering  into  more subgroups  found  larger  differences  between amputees than able-bodied subjects. Furthermore, a simple linear classifier was  able  to  discriminate between  able-bodied  and  amputee subjects   using   EMG   from   multiple   gestures   with 90\% accuracy.These   results   suggest that using   able-bodied subject   data   alone   may   be   insufficient   to   capture   the necessary  inter-subject  variance  when designing cross-user myoelectric control systems for prosthesis control.
\end{abstract}


\section{Introduction}
Although many applications of myoelectric control have been proposed in the literature since the 1990s, prosthesis control may still be considered as the predominant, and only commercial, application \cite{oskoei2007myoelectric}. Nevertheless, despite many laboratory-based advances in pattern recognition-based myoelectric control (>90\% classification accuracy) \cite{oskoei2007myoelectric}, myoelectric-controlled prostheses still make a relatively limited clinical and commercial impact (e.g., only a quarter of patients with upper extremity amputations chose to use a myoelectric prosthesis \cite{wright1995prosthetic}). This may be due to a gap between the academic state-of-the-art in myoelectric control and industry, which has been acknowledged and highlighted within the academic community \cite{campbell2020current,jiang2012myoelectric,scheme2011electromyogram}. One major limitation is high inter-subject variability, which limits the generalization of findings and necessitates frequent user-specific training and custom calibration \cite{kyranou2018causes,phinyomark2013feasibility}.

The main assumption of pattern recognition-based myoelectric control is that different types of muscle contractions exhibit distinguishable and repeatable signal patterns. Although distinguishable activation patterns are routinely found within a single user, there remain large differences between subjects. Most research studies, therefore, have adopted single-user (or subject-dependent) classification models, i.e., every user must train a system before his/her gestures can be recognized \cite{oskoei2007myoelectric}. Few studies have investigated cross-user (or subject-independent) models and results have shown a marked decrease from the state-of-the-art (from >90\% to 40\%-60\%) \cite{saponas2008demonstrating,kim2015real}. Moreover, due to difficulties with access to persons with upper extremity limb deficiencies, most research studies have developed and investigated pattern recognition-based myoelectric control systems using intact-limbed subjects. Although relatively consistent algorithmic trends exist between the intact-limbed and amputee populations, an overall decrease in performance has typically been reported for the latter \cite{scheme2011electromyogram,campbell2019differences}.

In order to facilitate the development of cross-user models, particularly for clinical applications of myoelectric control, more information about subject-related differences in electromyography (EMG) patterns is required. The purpose of this preliminary study was, therefore, to examine the differences in surface EMG patterns between intact-limbed and amputee subjects across a large set of hand and finger gestures. Results are explored using data visualization and cluster analysis techniques.

\section{Methods}

\subsection{EMG Data and Pre-Processing}
Surface EMG data used in this study were taken from two NinaPro (Non-Invasive Adaptive Prosthetics) databases (3 and 7) \cite{atzori2014electromyography,krasoulis2017improved}, which include data acquired from 20 intact-limbed subjects and 10 trans-radial amputated subjects. Additional details about the nature of the amputee subject data are shown in Table 1.

In these data sets, subjects performed a series of motions, including various individual-finger, hand, wrist, grasping, and functional movements. Databases 3 and 7 contain 52 and 40 total gestures, respectively, but the 38 common motions between the two databases were used for the present study. Each motion lasted 5 s, interrupted by 3-s rest time, and was repeated six times. Surface EMG data were collected using twelve Delsys Trigno Wireless electrodes; eight electrodes were equally spaced around the forearm (at the height of the radio-humeral joint), two electrodes were placed on the flexor and extensor digitorum superficialis muscles, and the remaining two electrodes were placed on the biceps and triceps brachii muscles. The sampling frequency was set to 2000 Hz. The data were cleaned of 50 Hz (and its harmonics) power-line interference using a Hampel filter. Erroneous movement labels were corrected by applying a generalized likelihood ratio algorithm \cite{atzori2014electromyography}.

\begin{table}[h]
    \centering
    \caption{Clinical characteristics of the amputee subjects (A1 and A2 from NinaPro Database 7 and A3-A10 from NinaPro Database 3). ‘n/a’ denotes data not available.}
    \label{tab:Amp_Data}
    \begin{tabular}{c c c c c}
    \toprule
        Subject & Amputated Hand & Years Since Amputation & Remaining Forearm (\%) & Cause of Amputation \\\midrule
        A1  & Right         & 6  & n/a & Accident \\
        A2  & Right         & 18 & n/a & Cancer   \\
        A3  & Left          & 6  & 70  & Accident \\
        A4  & Right         & 5  & 30  & Accident \\
        A5  & Right \& Left & 1  & 40  & Accident \\
        A6  & Right         & 7  & 0   & Accident \\
        A7  & Right         & 5  & 50  & Accident \\
        A8  & Right         & 14 & 90  & Accident \\
        A9  & Right         & 2  & 50  & Accident \\
        A10 & Right         & 5  & 90  & Cancer   \\\bottomrule
    \end{tabular}
    
\end{table}

\section{Processing and Evaluation}
The pre-processed EMG data were segmented for feature extraction using a window size of 200 ms and an increment of 100 ms. The commonly used Hudgins’ time domain features \cite{hudgins1993new}; mean absolute value (MAV), waveform length (WL), zero crossing (ZC), and slope sign change (SSC), were extracted from each window. A feature vector was then created from a series of the overlapped windows for further analyses.

Hierarchical cluster analysis (HCA) was used to create a dendrogram that identified homogeneous myoelectric patterns across the entire participant group (30 subjects). Briefly, HCA builds a hierarchical tree by combining a pair of clusters that leads to the minimum increase in total within-cluster variance after merging (Ward’s criterion \cite{ward1963hierarchical}), where the increase is a weighted squared Euclidean distance between cluster centers. Subjects in the same group have higher similarity (on average across 38 gestures, 12 muscles, and 6 repetitions) than the subjects in the other groups. Clusters in the data are determined by considering the height (or the distance) of each link in the cluster tree compared to the heights of the lower level links in the tree. If a link has a small increase in the height relative to the links below, it means that there are less distinct patterns differentiating the subjects joined at that level. Conversely, if a link height significantly differs from the links below, it means that there are more distinct patterns between them. This measure is referred to as the inconsistency coefficient.

Data visualization using principal component analysis (PCA), a commonly used feature projection method, was performed to better understand these complex myoelectric patterns. The main purpose of PCA is to summarize the important variance information in the data into the first few principal components (PCs), to facilitate visualization of distance and relatedness between populations in a reduced dimension. The identified PCs are linear combinations of the original features that can be used to express the data in a reduced form.

Finally, classification accuracies were computed using a linear discriminant analysis (LDA) classifier and a leave-one-out cross-validation technique to measure the performance of classification models in discriminating between gestures and between subjects. For gesture recognition, six clinically relevant motions were evaluated: wrist flexion, wrist extension, forearm pronation, forearm supination, power grip, and pinch grip. Classical within-subject gesture recognition was performed using leave-one-repetition-out cross-validation. For subject recognition, overall signal patterns were used (combining features from all repetitions of motions) in a leave-one-subject-out cross-validation approach. The goal of this task was to evaluate whether data could be classified as being from an able-bodied or amputee subject. This classification task was also repeated using each individual 200ms window of EMG data, again in a leave-one-subject-out cross-validation. 

\section{Results}
To validate previously reported results for intact-limbed and amputee subjects, the conventional gesture classification performance was computed for each group (Fig. 1). In keeping with previous findings, classification accuracies for the group of 20 intact-limbed subjects were significantly higher than the group 10 amputees (90.54\% $\pm$ 3.6\% $>$ 80.58\% $\pm$ 9.8\%; \textit{p} $<$ 0.01).

The results of the subject cluster analysis are shown in Fig. 2. It can be seen that the difference between the height of the links that connect the clusters (amputee and intact-limbed groups) and the mean height of the two links directly below is largest. In addition, the differences between the height of the links decreased as the number of clusters increased, and a plateau was found after six clusters were created. Thus, in this study, the two-cluster and the six-cluster solutions were employed.

\begin{figure}
    \centering
    \includegraphics[width=0.6\textwidth]{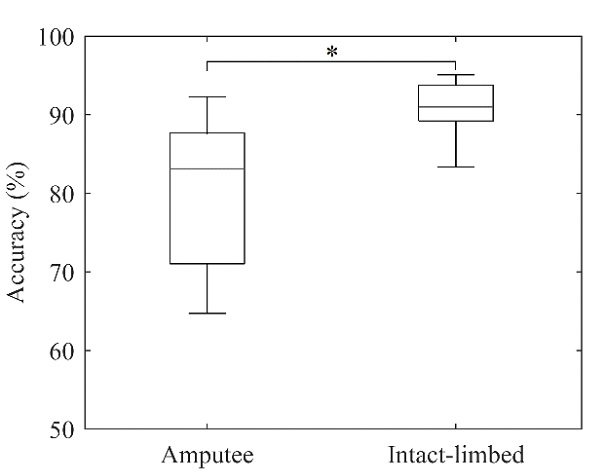}
    \caption{Figure 1: Box plot of gesture classification accuracies using an LDA classifier with Hudgins’ time domain features for amputee and intact-limbed subjects. * indicates significant difference (\textit{p} $<$ 0.01).}
    \label{fig:my_label}
\end{figure}

\begin{figure}
    \centering
    \includegraphics[width=0.6\textwidth]{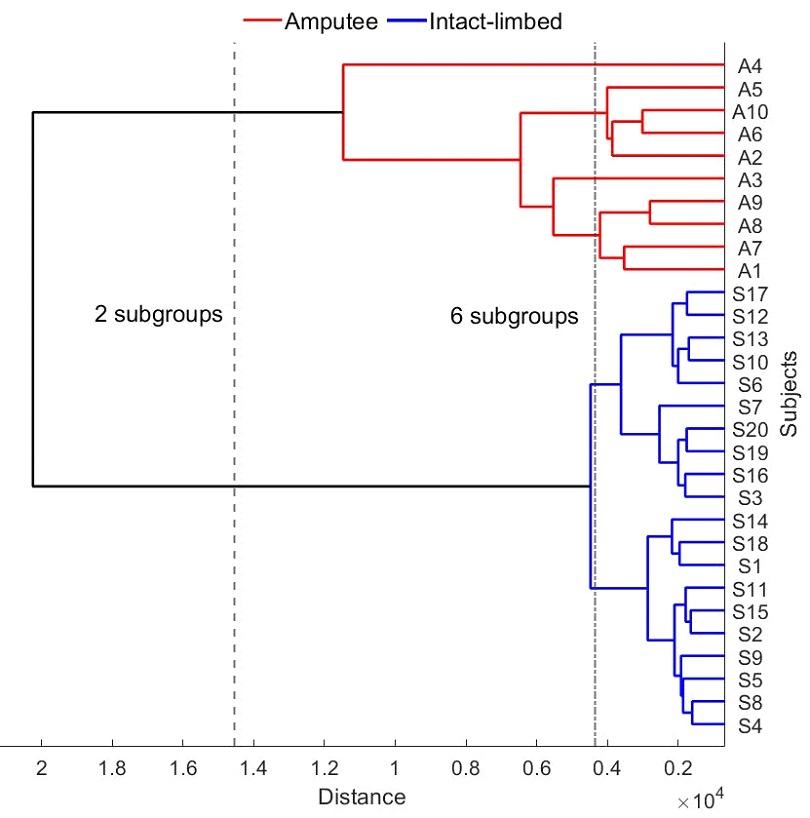}
    \caption{Ward’s linkage dendrogram of the hierarchical clustering of the overall myoelectric patterns representing the two-group and the six-group solutions. Participant numbers are indicated.}
    \label{fig:my_label}
\end{figure}

\begin{figure}
    \centering
    \includegraphics[width=0.6\textwidth]{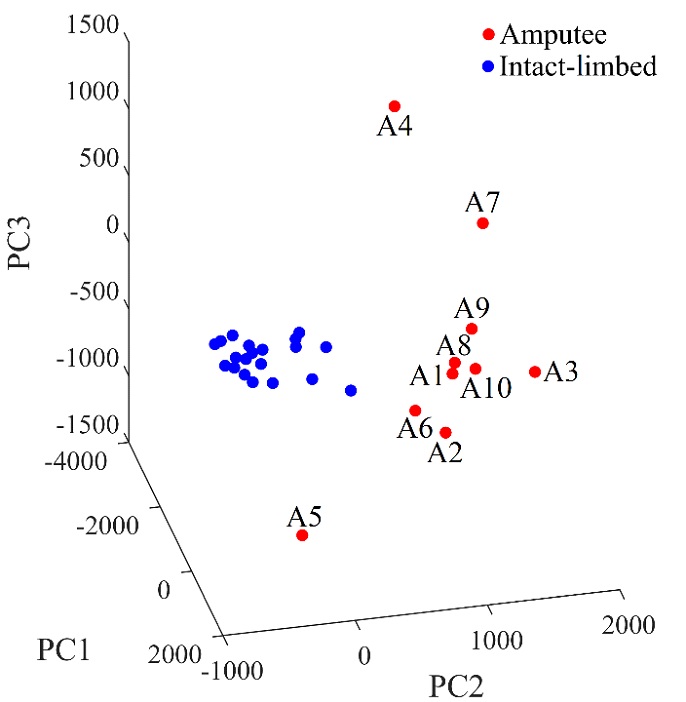}
    \caption{Scatter plot of the first three PCs representing overall myoelectric patterns for 10 amputees (red dot) and 20 intact-limbed subjects (blue dot).  The first three PCs explained 54\% of the total variance.}
    \label{fig:my_label}
\end{figure}

When partitioning into two clusters (at the leftmost vertical dotted line in Fig. 2), Cluster 1 was found to consist purely of the amputee subjects (A1-A10) and Cluster 2, of purely intact-limbed subjects (S1-S20). When partitioning into six clusters (at the rightmost vertical line in Fig. 2), the previous clusters were retained, but were further subdivided. Cluster 1 was partitioned into 4 subgroups, with 1 subject in Cluster 1A, 4 subjects in Cluster 1B, 1 subject in Cluster 1C, and 4 subjects in Cluster 1D. The previous Cluster 2 was partitioned into 2 subgroups, with 10 subjects in Cluster 2A and 10 subjects in Cluster 2B.

Fig. 3 shows the projection of all subjects into PCA space. Two distinct clusters of patterns can be seen, highlighting the differences between intact-limbed and amputee subjects. A classification accuracy of 90\% was found when using a simple LDA classifier to classifier whether the data from a given subject was able-bodied or amputee based on their overall signal patterns. Although overall signal patterns were distinct, no differences were observable between the groups even when classifying a single frame of EMG as being from an able-bodied or amputee subject. A mean accuracy of 66\% (min: 46\%, max: 78\%, chance: 66\%) was observed across all subjects and motions classes.

\section{Discussion}

The main purpose of this study was to determine whether myoelectric patterns for intact-limbed and amputee subjects could be classified into homogeneous subgroups. The HCA approach was successful in identifying two distinct subgroups (yielding the highest inconsistency coefficient value: 4.38) based on overall myoelectric patterns. Although it would be expected that there are differences between intact-limbed and amputee subjects, it is quite surprising that an unsupervised learning algorithm could create two subgroups that discriminate myoelectric patterns of amputees and intact- limbed subjects nearly perfectly (Fig. 2 and 3). From observation of Fig. 3, it appears as though a non-linear classifier could achieve 100\% classification using only 2-3 PCs. Campbell et al. \cite{campbell2019differences} investigated the differences between amputees and intact-limbed subjects using 58 state-of-the-art myoelectric features and suggested that most features in both time domain and frequency domain extract the same information for both subject groups. However, the migration of several amputee EMG features was found and can partially explain the performance degradation in amputee subjects (Fig. 1) (i.e., less information content is extracted using some EMG features for amputees). These findings suggest that when access to amputee populations is limited and able-bodied data is supplemented, outcomes of investigations on EMG features, dimensionality reduction, and classification algorithms should expect performance degradation when translating back to amputee populations. If a research study would like to develop a cross-user or subject-independent classification model for myoelectric-controlled prostheses, EMG data from amputee subjects is likely necessary given the noticeable difference in their patterns as compared to their intact-limbed counterparts (Fig. 1 and 3).

When 3-5 clusters were formed in Fig. 2, one group consisting of all the intact-limbed subjects remained consistent while the amputee group was partitioned into subgroups. This finding suggests that inter-subject variability in the amputee population is higher than between able-bodied subjects. A higher standard deviation of the classification accuracies for amputees (Fig. 1) also supported a higher inter-subject variability in amputee population. When the number of clusters was increased to six (yielding the second highest inconsistency coefficient value: 3.11), intact-limbed subjects were also divided into two subgroups. Some interesting characteristics of the six subgroups of subjects were found. For the two able-bodied subgroups, Cluster 2A provided slightly higher feature values compared to Cluster 2B. Cluster 1A, which contained only subject A4, provided the highest values for amplitude-based features (MAV and WL) among all subgroups but provided the lowest values for the complexity and frequency information-based features (ZC and SSC). It should be noted that the variance of feature values for this subject was very high, which could be due to noise or poor contraction repeatability. 

Cluster 1D, which consists of 4 amputee subjects, provided the lowest values for the amplitude-based features, but the highest values for the complexity and frequency information-based features. It should be noted that most subjects in this group had prior experience in using a myoelectric prosthesis, suggesting that learning may play a role in cross-user differences. Cluster C1 consisted of only subject A3, the only subject with a left amputated hand and using a cosmetic prosthesis. Both subjects with an amputation due to cancer, were clustered together, in Cluster 1B. No meaningful trends were found for other clinical characteristics such as years since amputation, the remaining forearm percentage, degree of phantom limb sensation, and DASH (disability of the arm, shoulder and hand) score.

Overall, these findings suggest that the adoption of data from able-bodied subjects for the investigation of EMG features, dimensionality reduction, and classification algorithms, should be done with caution when focused on clinical applications for amputees. Specifically, even unsupervised clustering methods identified two distinct groups of subjects: one with all amputees and the other with all intact-limbed subjects. Of the subgroups, the amputee subgroup demonstrated much higher inter-subject variability. These results suggest that EMG data from amputee subjects is necessary for creating cross-user myoelectric-controlled prostheses, as their myoelectric patterns are considerably different than their intact-limbed counterparts.

\bibliographystyle{unsrt}  

\bibliography{references}

\begin{thebibliography}{10}

\bibitem{oskoei2007myoelectric}
Mohammadreza~Asghari Oskoei and Huosheng Hu.
\newblock Myoelectric control systems—a survey.
\newblock {\em Biomedical signal processing and control}, 2(4):275--294, 2007.

\bibitem{wright1995prosthetic}
Thomas~W Wright, Arlene~D Hagen, and Michael~B Wood.
\newblock Prosthetic usage in major upper extremity amputations.
\newblock {\em Journal of Hand Surgery}, 20(4):619--622, 1995.

\bibitem{campbell2020current}
Evan Campbell, Angkoon Phinyomark, and Erik Scheme.
\newblock Current trends and confounding factors in myoelectric control: Limb
  position and contraction intensity.
\newblock 2020.

\bibitem{jiang2012myoelectric}
Ning Jiang, Strahinja Dosen, Klaus-Robert Muller, and Dario Farina.
\newblock Myoelectric control of artificial limbs—is there a need to change
  focus?
\newblock {\em IEEE Signal Processing Magazine}, 29(5):152--150, 2012.

\bibitem{scheme2011electromyogram}
Erik Scheme and Kevin Englehart.
\newblock Electromyogram pattern recognition for control of powered upper-limb
  prostheses: state of the art and challenges for clinical use.
\newblock {\em Journal of Rehabilitation Research \& Development}, 48(6), 2011.

\bibitem{kyranou2018causes}
Iris Kyranou, Sethu Vijayakumar, and Mustafa~Suphi Erden.
\newblock Causes of performance degradation in non-invasive electromyographic
  pattern recognition in upper limb prostheses.
\newblock {\em Frontiers in neurorobotics}, 12:58, 2018.

\bibitem{phinyomark2013feasibility}
Angkoon Phinyomark, Franck Quaine, Sylvie Charbonnier, Christine Serviere,
  Franck Tarpin-Bernard, and Yann Laurillau.
\newblock A feasibility study on the use of anthropometric variables to make
  muscle--computer interface more practical.
\newblock {\em Engineering Applications of Artificial Intelligence},
  26(7):1681--1688, 2013.

\bibitem{saponas2008demonstrating}
T~Scott Saponas, Desney~S Tan, Dan Morris, and Ravin Balakrishnan.
\newblock Demonstrating the feasibility of using forearm electromyography for
  muscle-computer interfaces.
\newblock In {\em Proceedings of the SIGCHI Conference on Human Factors in
  Computing Systems}, pages 515--524, 2008.

\bibitem{kim2015real}
Jongin Kim, Dongrae Cho, Kwang~Jin Lee, and Boreom Lee.
\newblock A real-time pinch-to-zoom motion detection by means of a surface
  emg-based human-computer interface.
\newblock {\em Sensors}, 15(1):394--407, 2015.

\bibitem{campbell2019differences}
Evan Campbell, Angkoon Phinyomark, Ali~H Al-Timemy, Rami~N Khushaba, Giovanni
  Petri, and Erik Scheme.
\newblock Differences in emg feature space between able-bodied and amputee
  subjects for myoelectric control.
\newblock In {\em 2019 9th International IEEE/EMBS Conference on Neural
  Engineering (NER)}, pages 33--36. IEEE, 2019.

\bibitem{atzori2014electromyography}
Manfredo Atzori, Arjan Gijsberts, Claudio Castellini, Barbara Caputo,
  Anne-Gabrielle~Mittaz Hager, Simone Elsig, Giorgio Giatsidis, Franco
  Bassetto, and Henning M{\"u}ller.
\newblock Electromyography data for non-invasive naturally-controlled robotic
  hand prostheses.
\newblock {\em Scientific data}, 1(1):1--13, 2014.

\bibitem{krasoulis2017improved}
Agamemnon Krasoulis, Iris Kyranou, Mustapha~Suphi Erden, Kianoush Nazarpour,
  and Sethu Vijayakumar.
\newblock Improved prosthetic hand control with concurrent use of myoelectric
  and inertial measurements.
\newblock {\em Journal of neuroengineering and rehabilitation}, 14(1):71, 2017.

\bibitem{hudgins1993new}
Bernard Hudgins, Philip Parker, and Robert~N Scott.
\newblock A new strategy for multifunction myoelectric control.
\newblock {\em IEEE Transactions on Biomedical Engineering}, 40(1):82--94,
  1993.

\bibitem{ward1963hierarchical}
Joe~H Ward~Jr.
\newblock Hierarchical grouping to optimize an objective function.
\newblock {\em Journal of the American statistical association},
  58(301):236--244, 1963.

\end{thebibliography}

\end{document}